\documentclass[twocolumn]{pasj02}
\usepackage[switch, mathlines]{lineno} 
\usepackage{natbib}
\usepackage{threeparttable}
\usepackage{xcolor}
\definecolor{xlinkcolor}{rgb}{0, 0, 1}
\usepackage[bookmarks=true, pdfnewwindow=true, colorlinks=true, linkcolor=xlinkcolor, citecolor=xlinkcolor, filecolor=xlinkcolor, urlcolor=xlinkcolor, final=true]{hyperref}
\usepackage{bookmark}

\jyear{2025}
\Received{}
\Accepted{}

\graphicspath{{./}} 

\begin{document} 

\title{Chemical enrichment in the Ophiuchus cluster core studied by high-resolution XRISM spectroscopy}

\author{
Kotaro~\textsc{Fukushima},\altaffilmark{1}\altemailmark\orcid{0000-0001-8055-7113} \email{kxfukushima@gmail.com} 
Yutaka~\textsc{Fujita},\altaffilmark{2}\orcid{0000-0003-0058-9719}
Kosuke~\textsc{Sato},\altaffilmark{3}\orcid{0000-0001-5774-1633}
Yasushi~\textsc{Fukazawa},\altaffilmark{4}\orcid{0000-0002-0921-8837}
and Marie~\textsc{Kondo}\altaffilmark{5}\orcid{0009-0005-5685-1562}
}
\altaffiltext{1}{Institute of Space and Astronautical Science, JAXA, 3-1-1 Yoshinodai, Chuo-ku,
Sagamihara, Kanagawa 252-5210, Japan}
\altaffiltext{2}{Department of Physics, Tokyo Metropolitan University, 1-1 Minami-Osawa, Hachioji, Tokyo 192-0397, Japan}
\altaffiltext{3}{Department of Astrophysics and Atmospheric Sciences, Kyoto Sangyo University, Motoyama, Kamigamo, Kita-ku, Kyoto, Kyoto 603-8555, Japan}
\altaffiltext{4}{Department of Physics, Hiroshima University, 1-3-1 Kagamiyama, Higashi-Hiroshima, Hiroshima 739-8526, Japan}
\altaffiltext{5}{Department of Physics, Saitama University, 255 Shimo-Okubo, Sakura-ku, Saitama, Saitama 338-8570, Japan}



\maketitle

\begin{abstract}
Galaxy clusters provide an ideal laboratory for investigating the chemical enrichment history of the universe
because they host the hot intracluster medium (ICM), which contains various chemical elements.
The X-ray observations have constituted a unique way to measure the element abundance
and composition of the ICM due to their prominent emission lines in the $0.1$--$10$\,keV range.
We explore the metal abundances and chemical enrichment in the cool-core galaxy cluster, Ophiuchus,
by using a $217$\,ks XRISM data set.
The abundances of Si, S, Ar, Ca, Cr, Mn, Fe, and Ni
are accurately determined using high-resolution spectroscopy.
We find that the average uncertainties of chemical composition,
which are reported as X/Fe ratios, are only 10--20\%.
The X/Fe abundance pattern of the Ophiuchus centre is remarkably consistent with solar,
which is reminiscent of the Hitomi constraint on the Perseus core.
The observed abundance pattern can be replicated globally by linear combination models
of core-collapse, including massive progenitors, and Type~Ia supernovae.
While nucleosynthesis models typically underestimate the Ca/Fe ratio,
a substantial contribution of Ca-rich gap transients may help improve the deficit of Ca.
High-resolution spectroscopic data can enable us to estimate the underlying impact
on the chemical enrichment from subclasses of Type~Ia supernovae.
\end{abstract}


\section{Introduction \label{sec:intro}}

Supernova (SN) explosions play a crucial role in the synthesis and ejection of metals in the universe.
Generally, core-collapse SNe (CCSNe) produce light $\alpha$-elements of O, Ne, and Mg \citep[e.g.,][]{Nomoto13},
and Type Ia SNe (SNe~Ia) dominate the forge of the Fe-peak elements (Cr, Mn, Fe, and Ni; \citealt{Seitenzahl13b}).
A significant fraction of these elements now resides within the intracluster medium (ICM)
that permeates the entire volume of galaxy clusters.
Thus, measuring the metal abundances in the hot gas in galaxy clusters
gives us a unique understanding of the evolution of chemical enrichment on the most enormous scale in the universe.

The elemental abundances in the ICM have been well investigated
using X-ray spectroscopic data because prominent emission lines of metals
are present in the $1$--$10$\,keV band \citep[][for recent reviews]{Mernier18c, Sanders23}.
Thanks to the fleet of modern X-ray observatories (Chandra, XMM-Newton, Suzaku, and Hitomi),
we have made considerable progress in studies on the cosmic abundances
for O, Ne, Mg, Si, S, Ar, Ca, Fe, and Ni, especially around the brightest cluster galaxies (BCGs).
These metals are considered primary products of SNe in the BCG.
We can also test the model calculations of the SN nucleosynthesis
by comparing them with the observed abundance pattern \citep[e.g.,][]{Sato07, Mernier16b}.
In particular, the Hitomi observation of the Perseus cluster revealed
that the pattern of abundance ratio to Fe (X/Fe) in the ICM
is entirely consistent with our solar \citep[][]{Simionescu19}.
Hence, whatever the meaning of this agreement with the solar composition,
constructing a chemical enrichment model that can explain the abundance pattern measured by X-ray observations
will directly provide a key to unravelling the chemical enrichment history of our local environment.

The recent X-ray observatory XRISM has been in operation since 2023 \citep{Tashiro25},
installing the X-ray telescope with a micro-calorimeter instrument offering a high-energy resolution
of about $5$\,eV in the $1.7$--$12$\,keV band (Resolve, \citealt{Ishisaki25, Kelley25}).
As for now, XRISM has demonstrated its great suitability for measuring the bulk and turbulent velocities of the ICM
\citep[e.g.,][]{Rose25, XRISM25e, XRISM25c, XRISM25d, XRISM25a}.
In addition, the non-dispersive high-resolution spectra provided by calorimetric spectroscopy
are also a valuable asset for abundance measurements in the ICM \citep[][]{Hitomi17, Simionescu19}.
Therefore, our next leap will be achieved by XRISM.

In this sense, the Ophiuchus cluster (redshift\,$= 0.00296$, \citealt{Durret15})
is an excellent target as the second X-ray-brightest cluster to Perseus.
Despite the presence of a cool core, Ophiuchus hosts a relatively high-temperature ICM,
$kT \sim 7$--$9$\,keV at the innermost region \citep{Fujita08},
resulting in a simple spectrum dominated by Fe~He$\alpha$ and Ly$\alpha$ lines.
This advantage enables us to easily model the ICM emission
compared to typical cool-core systems, such as Centaurus,
considerably contaminated by Fe~\textsc{xxiv} or \textsc{xxiii} lines.
The Fe abundance at a central $100$\,kpc region is moderate about $0.6$\,solar
\citep[][]{Fujita08, Million10a, Werner16, Gatuzz23c}.
\citet{Gatuzz23c} has also reported that the Si, S, Ar, and Ca abundances
can be constrained using the stacked XMM-Newton data
while only the upper limit has been obtained for Ni.

In this paper, we will analyse the Resolve data of the central region of the Ophiuchus galaxy cluster.
We focused on the abundances of Si, S, Ar, Ca, Fe, and Ni throughout the entire core.
Recent SN nucleosynthesis yields from some literature are examined
to reproduce the observed abundances obtained with Resolve.
This paper is structured as follows. In Section\,\ref{sec:obs}, we summarise our XRISM observation and data reduction.
In Section\,\ref{sec:result}, details of our spectral analysis are demonstrated.
We interpret and discuss the results in Section\,\ref{sec:discuss}.
In this work, we adopt the cosmological parameters of
$H_0=70$\,km\,s$^{-1}$\,Mpc$^{-1}$, $\Omega_m=0.3$, and $\Omega_{\Lambda}=0.7$.
All abundances are expressed relative to the proto-solar values from \citet{Lodders09}.
The statistical errors are given at the $1 \sigma$ confidence level unless stated otherwise.

\section{Observation and Data reduction \label{sec:obs}}

\begin{figure}
\begin{center}

\includegraphics[width=\columnwidth]{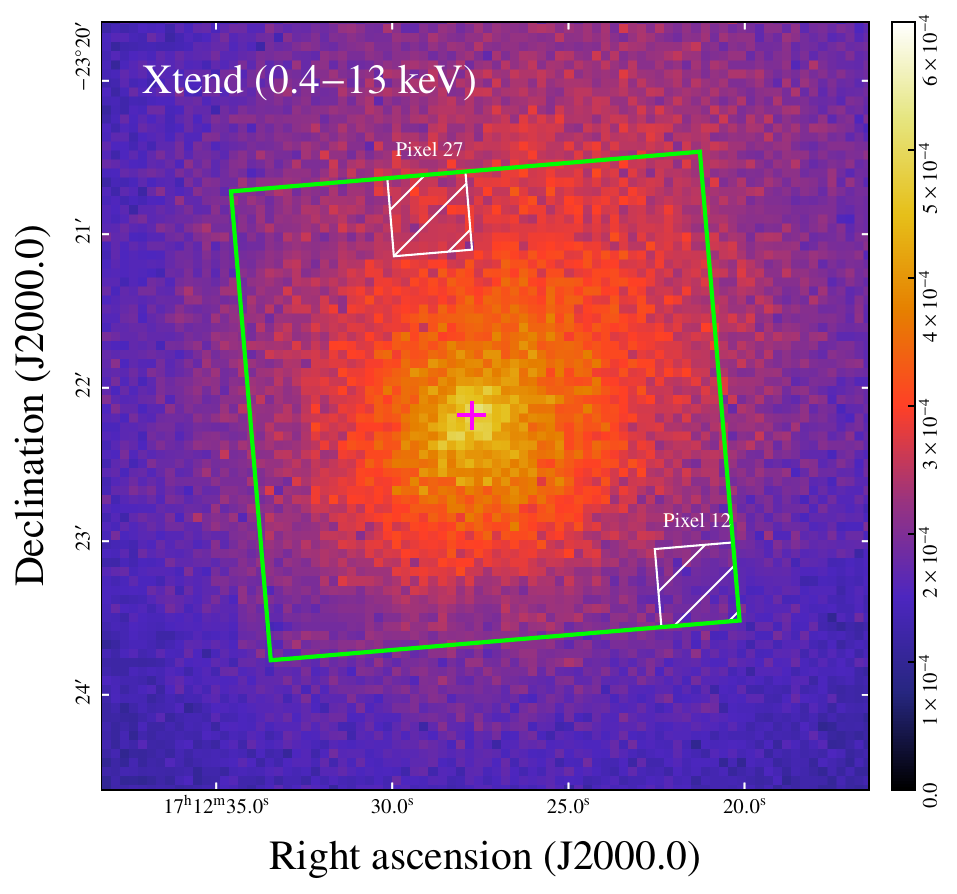}

\end{center}

\caption{Xtend exposure-corrected image in the $0.4$--$13$\,keV band of the Ophiuchus cluster.
Colour bar indicates the flux in units of count\,s$^{-1}$.
The larger square indicates the Resolve FoV.
The white-hatched ones represent the excluded pixel channels, 12 and 27.
The cross marks the position of the BCG.
\label{fig:img}}

\end{figure}

The central region of the Ophiucus cluster was observed by XRISM
from March 31 to April 6, 2025 ($\texttt{OBSID} = \texttt{201006010}$),
aiming to (R.A., delc.)\,$= (258.^\circ115, -23.^\circ369)$.
Figure\,\ref{fig:img} shows the X-ray image of the Ophiuchus centre observed
by the X-ray CCD array onboard XRISM (Xtend, \citealt{Noda25, Uchida25}).
In this work, we only use the data from Resolve, a microcalorimeter instrument
covering a $3.1 \times 3.1$\,arcmin$^2$ field of view (FoV) with 36 pixels \citep[][]{Ishisaki25, Kelley25}.
In the closed gate valve configuration, the sensitive energy band of Resolve spans about $1.7$--$12$\,keV.
The Resolve data were processed with the latest version 3 pipeline software
($\texttt{PROCVER} = \texttt{03.00.013.010}$) and analysed using XRISM \texttt{FTOOLS}
wrapped in HEASoft version 6.35 and calibration database version 11
(XRISM \texttt{CALDB}; \texttt{gen20241115} and \texttt{rsl20250315}).
Following the standard screening procedures described in the XRISM Data Reduction Guide
\footnote{$\langle$\url{https://heasarc.gsfc.nasa.gov/docs/xrism/analysis/abc\_guide/xrism\_abc.html}$\rangle$},
we extracted high-resolution events from the entire array shown in figure\,\ref{fig:img},
which constructs a spectrum with an energy resolution of $5$\,eV at $6$\,keV averaged over FoV.
The cleaned exposure time of the Resolve event is $217$\,ks after the screening.
We generated a response matrix using the \texttt{rslrmf} task with $\texttt{size} = \texttt{X}$
\footnote{With this option, the Gaussian core, the exponential tail, the escape peaks, the Si~K$\alpha$ emission line,
and the electron loss continuum are taken into account for a line spread function.}
and simulated an effective area with the \texttt{xrtraytrace} and \texttt{xaxmaarfgen} procedures.
We excluded pixel channels 12 and 27 from this work following the prescription of \citet{Fujita25}.

\section{Analysis and Results \label{sec:result}}

\begin{figure}
\begin{center}

\includegraphics[width=\columnwidth]{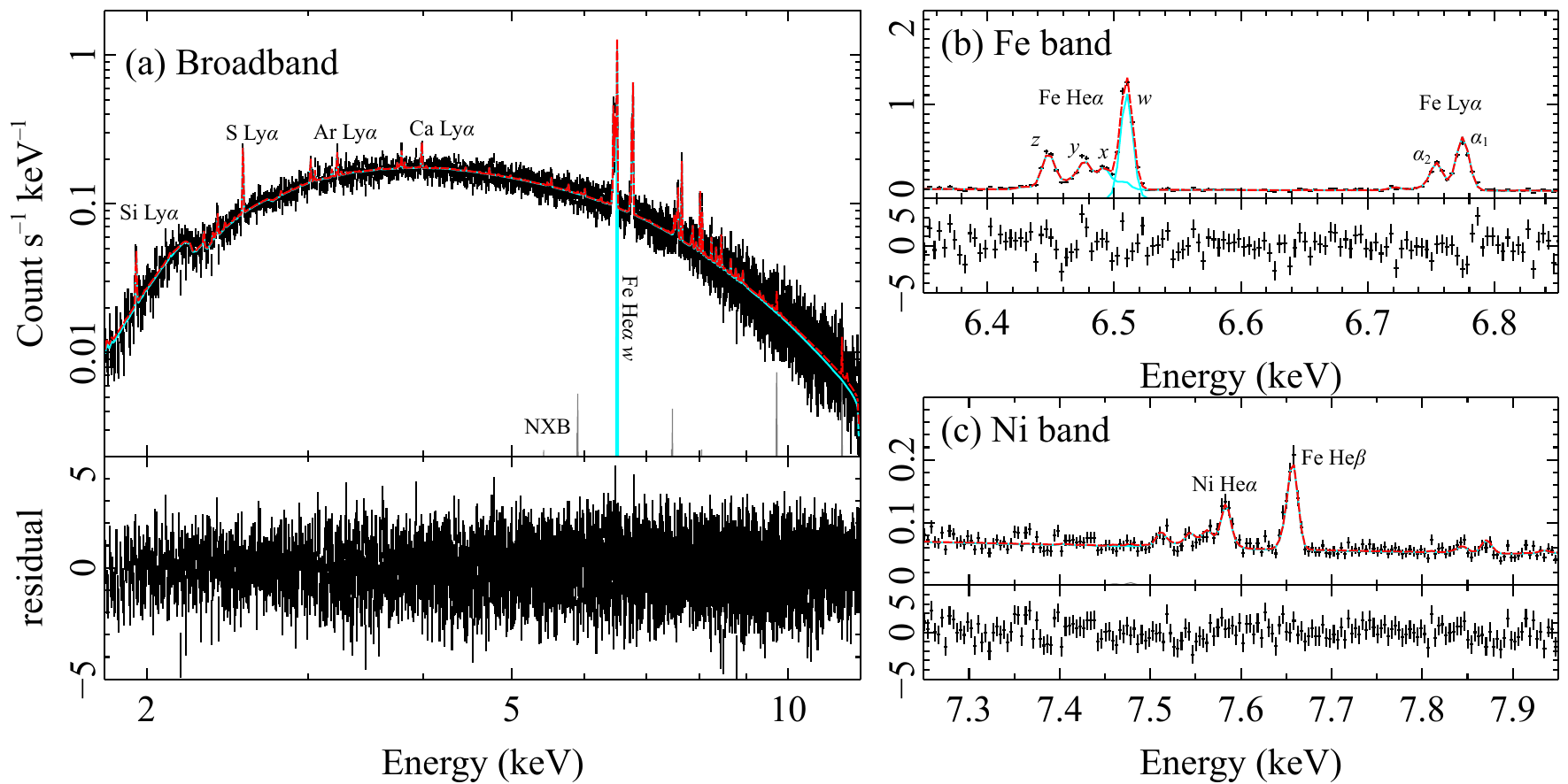}

\end{center}

\caption{(a) Broadband Resolve spectrum of the Ophiuchus centre.
The best-fitting model is shown by a red dashed line, which comprises an ICM component
with a Gaussian for the Fe~He$\alpha$-$w$ line shown by cyan solid lines.
A thin grey line indicates the NXB component.
Residuals interms of $(\textup{data} - \textup{model})/\sqrt{\textup{model}}$ are also shown.
(b, c) Zoom-up views around Fe and Ni lines.
The prominent metal emission lines we focus on in this work are marked in each panel.
\label{fig:spec}}

\end{figure}

\begin{table*}[htbp]
\begin{threeparttable}

\caption{The best-fitting spectral parameters of the Ophiuchus centre for each modelling. \label{tab:fovpars}}

\centering

\begin{tabular}{@{}lc@{\:\:}c@{\:\:}c@{\:\:}c@{\:\:}c@{\:\:}c}\\ \hline\hline
 & \multicolumn{2}{c}{1T} & \multicolumn{2}{c}{2T} & \multicolumn{2}{c}{DEM\tnote{$*$}} \\ \hline
 \multicolumn{7}{c}{temperature} \\
$kT_\textup{1}$ & \multicolumn{2}{c}{$7.21 \pm 0.06$} & \multicolumn{2}{c}{$11.0 \pm 0.6$} & \multicolumn{2}{c}{$-$} \\
$kT_\textup{2}$ & \multicolumn{2}{c}{$-$} & \multicolumn{2}{c}{$4.5 \pm 0.2$} & \multicolumn{2}{c}{$-$} \\
$kT_\textup{mean}$ & \multicolumn{2}{c}{$-$} & \multicolumn{2}{c}{$-$} & \multicolumn{2}{c}{$7.2 \pm 0.3$ ($7.22 \pm 0.19$)} \\
$\sigma_{kT}$ & \multicolumn{2}{c}{$-$} & \multicolumn{2}{c}{$-$} & \multicolumn{2}{c}{$5.1 \pm 0.8$ ($4.8 \pm 0.6$)} \\ \hline
\multicolumn{7}{c}{metal content} \\
 & global & local & global & local & global & local \\
Fe & $0.537 \pm 0.008$ & $0.521 \pm 0.011$ & $0.623 \pm 0.016$ & $0.59 \pm 0.04$ & $0.611 \pm 0.017$ ($0.617 \pm 0.013$) & $0.602 \pm 0.013$ ($0.622 \pm 0.018$) \\
Si/Fe & $1.4 \pm 0.3$ & $1.3 \pm 0.3$ & $0.97 \pm 0.19$ & $1.0 \pm 0.2$ & $0.81 \pm 0.16$ ($0.82 \pm 0.17$) & $0.84 \pm 0.17$ ($0.82 \pm 0.17$) \\
S/Fe & $1.83 \pm 0.15$ & $1.74 \pm 0.16$ & $1.30 \pm 0.11$ & $1.14 \pm 0.18$ & $1.21 \pm 0.10$ ($1.21 \pm 0.10$) & $1.17 \pm 0.10$  ($1.15 \pm 0.10$) \\
Ar/Fe & $1.22 \pm 0.19$ & $1.20 \pm 0.19$ & $0.90 \pm 0.18$ & $0.90 \pm 0.18$ & $0.91 \pm 0.14$ ($0.91 \pm 0.14$) & $0.94 \pm 0.15$  ($0.90 \pm 0.14$) \\
Ca/Fe & $1.32 \pm 0.16$ & $1.36 \pm 0.18$ & $1.06 \pm 0.12$ & $1.3 \pm 0.3$ & $1.13 \pm 0.13$ ($1.11 \pm 0.13$) & $1.13 \pm 0.15$  ($1.09 \pm 0.15$) \\
Cr/Fe & $1.3 \pm 0.4$ & $1.4 \pm 0.4$ & $1.1 \pm 0.3$ & $1.4 \pm 0.5$ & $1.3 \pm 0.3$ ($1.2 \pm 0.3$) & $1.2 \pm 0.4$  ($1.2 \pm 0.4$) \\
Mn/Fe & $0.9 \pm 0.4$ & $0.8 \pm 0.5$ & $1.1 \pm 0.4$ & $0.9 \pm 0.5$ & $1.2 \pm 0.5$ ($1.1 \pm 0.4$) & $0.9 \pm 0.5$  ($0.8 \pm 0.5$) \\
Ni/Fe & $1.33 \pm 0.11$ & $1.36 \pm 0.13$ & $1.34 \pm 0.12$ & $1.39 \pm 0.18$ & $1.33 \pm 0.12$ ($1.31 \pm 0.11$) & $1.33 \pm 0.14$  ($1.28 \pm 0.13$) \\ \hline
\multicolumn{7}{c}{statistics} \\
C-statistic/dof & \multicolumn{2}{c}{18259/18547} & \multicolumn{2}{c}{18181/18544} & \multicolumn{2}{c}{18193/18546 (18190/18544)} \\ \hline
\end{tabular}


\begin{tablenotes}
\item[$*$] Fitting results without the Fe~He$\alpha$-$w$ line are provided in parentheses.
\end{tablenotes}

\end{threeparttable}
\end{table*}

\subsection{Spectral fitting \label{subsec:fit}}

Here, we analyse X-ray spectra extracted from the FoV of Resolve.
We use {\sc xspec} package version 12.15.0 \citep{Arnaud96}
and the latest \textsc{AtomDB} version 3.1.3 to produce a collisional ionisation equilibrium (CIE) plasma.
The non-X-ray background (NXB) contribution to the Resolve spectrum \citep[][]{XRISM24c}
is taken into account by including the model components with a power law and multiple Gaussians for instrumental lines
\footnote{$\langle$\url{https://heasarc.gsfc.nasa.gov/docs/xrism/analysis/nxb/nxb\_spectral\_models.html}$\rangle$},
in our spectral fitting, rather than subtracting that.
The spectrum is rebinned to have a minimum of 1 count per energy bin,
and we fit it using the C-statistics \citep[][]{Cash79},
to estimate parameters and their errors without bias \citep{Kaastra17}.
Figure\,\ref{fig:spec}(a) shows the broadband spectrum of the ICM in the Ophiuchus core.
Prominent Ly$\alpha$ emission lines from Si, S, Ar, Ca, and Fe are detected.
Additionally, there are He$\alpha$ lines of Fe and Ni;
in particular, Ni~He$\alpha$ is separated from the Fe~Ly$\beta$ line (figures\,\ref{fig:spec}(b) and \ref{fig:spec}(c)),
both of which have been thoroughly blended in ICM spectra by CCD \citep[e.g.,][]{dePlaa13, Fukushima22}.

To reproduce the ICM emission, we use the CIE model modified by a Galactic extinction
presented as $\texttt{phabs} \times \texttt{CIE}$.
The absorption column density $N_\textup{H}$ is fixed at $3.34 \times 10^{21}$ \citep[][]{Willingale13},
which includes both H~\textsc{i} and H$_2$ effects considering the low Galactic latitude ($b \sim 9^\circ$).
The photoelectric absorption cross sections are retrieved from \citet{Verner96}.
Note that $N_\textup{H}$ is unimportant in our energy band
as the absorption with only H~\textsc{i} \citep[][]{HI4PI16} provides consistent results.
Since the ICM emission is very bright at the entire FoV,
we do not include any astrophysical emission from the front or backside of the ICM.
For example, we estimate the cosmic X-ray background (CXB) contribution
that can affect the temperature estimation in the Resolve band.
The brightness of the CXB emission is well established
as $S_\textup{CXB} \sim 11$\,cm$^{-2}$\,s$^{-1}$\,sr$^{-1}$\,keV$^{-1}$ at $1$\,keV \citep[][]{Cappelluti17}.
While we include an absorbed powerlaw component with $\Gamma = 1.45$ with this brightness,
its flux accounts for only 0.92\% of that of the ICM component in the $2$--$10$\,keV range.
The AGN activity of this cluster is also negligible \citep[e.g.,][]{Fujita25}.

\subsection{Metal abundances \label{subsec:abund}}

First, we test the isothermal CIE modelling, denoted as the 1T model, where $\texttt{CIE} = \texttt{bvvapec}$.
We let the ICM temperature ($kT$), turbulence velocity, redshift,
and the Si, S, Ar, Ca, Cr, Mn, Fe, and Ni abundances free to vary.
The abundances of elements lighter than O are set to $1$\,solar,
and those of metals from O to Al are linked to Fe
\footnote{Note that these species do not make prominent line signals in the $1.8$--$12$\,keV band.}.
The abundances of the other elements with atomic numbers lying between the species above
are linked to those of the nearest lower-number elements (e.g., $\textup{Ti} = \textup{Ca}$).
We also perform local or narrowband fits within the energy bands
around the most prominent lines of each element, excluding faint Si.
The obtained $kT$ and elemental abundances are shown in table\,\ref{tab:fovpars}.
The metal content of the ICM from 1T is uniformly larger than unity.
One caveat is that the isothermal assumption of the multiphase ICM
could lead to an underestimation of the Fe abundance \citep[e.g.,][]{Gastaldello21}.
As described in \citet{Fujita25}, the ICM of the Ophiuchus centre exhibits at least a biphasic state.

Next, we assume a multiphase ICM temperature structure with $\texttt{CIE} = \texttt{bvvapec} + \texttt{bvvapec}$
called the 2T model and $\texttt{CIE} = \texttt{bvvgadem}$ referred to as the differential emission measure (DEM) model.
The metal content and the line-of-sight bulk velocity
are shared between the two components in the 2T model, and the velocity dispersions are calculated separately.
As summarised in table\,\ref{tab:fovpars}, these models result in a higher Fe abundance than that of 1T;
thus, the X/Fe ratios are closer to the solar values.
Both global and local abundances are consistent with each other and those from the 2T model,
maintaining the solar composition of the ICM of the Ophiuchus core.
Although both 2T and DEM improve the fit statistics, the statistical errors of abundance measurements
are smaller with DEM than with 2T, especially for the local Ca/Fe estimate.
This schism would be due to the coupling of the intensities of the two components and their abundance,
with the reproduction of the Ca line being about evenly contributed by both.
Therefore, we prefer the abundances from DEM with reasonable multi-temperature modelling.

Finally, we remove the Fe~He$\alpha$-$w$ line from the \textsc{AtomDB} code
and add a Gaussian to model this line as $\texttt{CIE} = \texttt{bvvgadem} + \texttt{gauss}$.
This prescription is used to account for resonance scattering in the central region
that reduces the Fe~He$\alpha$-$w$ line intensity \citep[e.g.,][]{Hitomi18a}.
We employed this method only in the DEM model.
As discussed in \citet{Fujita25}, the optical depth of the $w$ line at the very core of Ophiuchus can reach about 1,
possibly leading to resonance-scattering effects.
Indeed, the $w/z$ ratio measured by Gaussians is 15\% lower than the CIE prediction.
The derived abundances are given in parentheses in the DEM column of table\,\ref{tab:fovpars},
which are globally consistent with the original modelling and the solar abundance ratios.
Afterwards, we use these values as benchmarks for this work.
In this benchmark result, our detection of Cr and Mn in the Ophiuchus core
is at the $3.3 \sigma$ and $1.7 \sigma$ significance, respectively, with respect to the zero metallicity.
Modelling uncertainties of abundance measurements are provided as $1\sigma$ deviations
for each trial with 2T and DEM: $\sigma_\textup{Si/Fe} = 0.09$, $\sigma_\textup{S/Fe} = 0.06$,
$\sigma_\textup{Ar/Fe} = 0.014$, $\sigma_\textup{Ca/Fe} = 0.08$, $\sigma_\textup{Cr/Fe} = 0.08$,
$\sigma_\textup{Mn/Fe} = 0.16$, $\sigma_\textup{Ni/Fe} = 0.04$.
The differences between \textsc{AtomDB} v3.1.3 and \textsc{spexact} v3.08.02 \citep[][]{Kaastra96, Kaastra25}
are only up to 2\% in abundance ratios, which is small compared to the modelling biases of up to 11\%.

\section{Discussion \label{sec:discuss}}

\begin{figure}
\begin{center}

\includegraphics[width=\columnwidth]{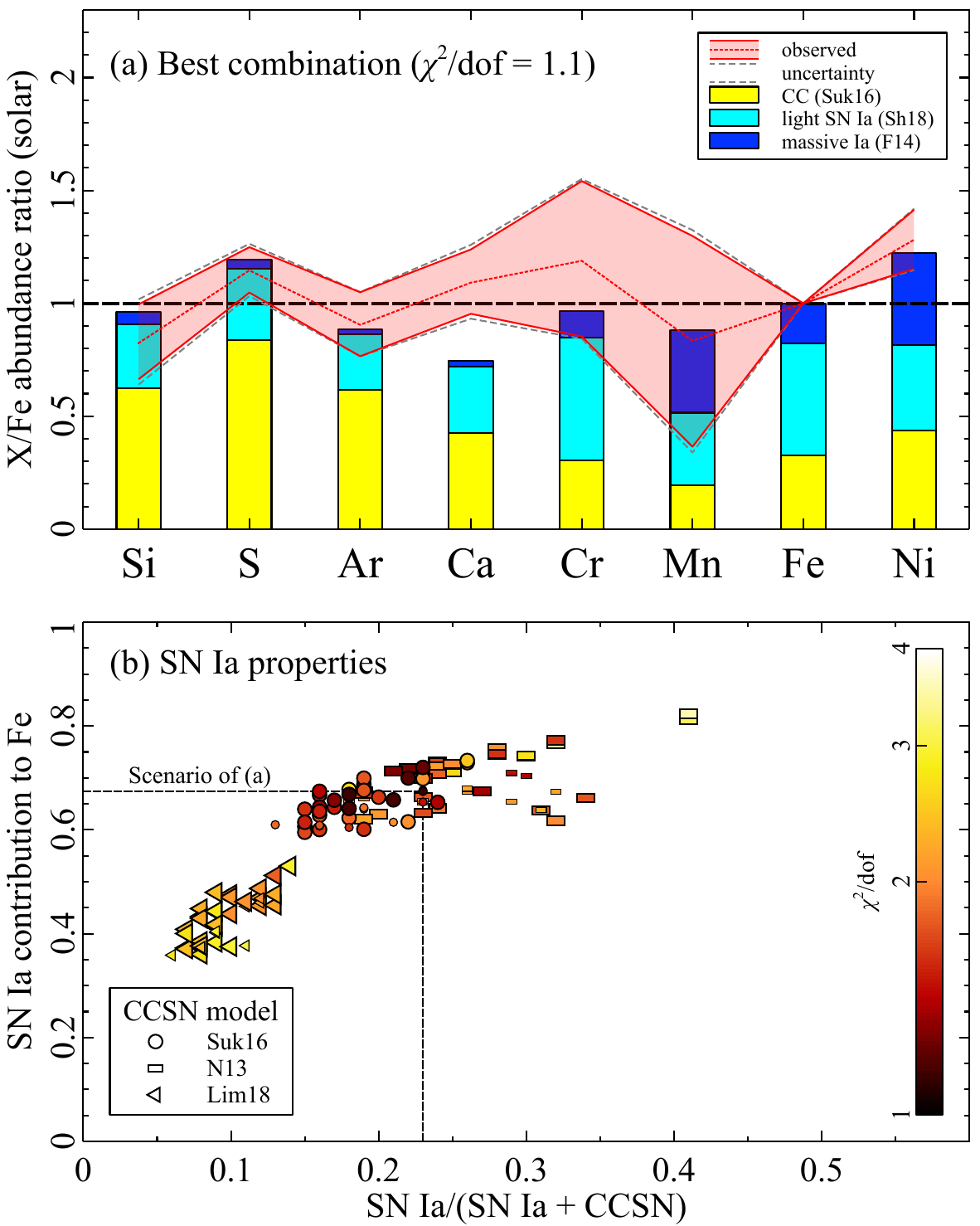}

\end{center}

\caption{(a) Observed relative abundance patterns obtained from FoV
and the best-fitting result of SN nucleosynthesis models comprising Suk16, F14, and Sh18.
The shaded area represents the observed pattern with statistical errors,
and the dashed lines limit their total $1 \sigma$ confidence levels
that are calculated by summing the statistical and modelling uncertainties
(see Section\,\ref{subsec:abund}) in quadrature.
The histogram shows contributions to abundance ratios from each SN type.
(b) SN~Ia contribution plotted against its number fraction to total SNe.
Colour scale indicates the $\chi^2$/dof values for each combination.
The best-fitting model demonstrated in (a) is marked by dashed lines.
The small markers show the models, including pure-deflagration options (see table\,\ref{tab:snmodel}).
\label{fig:patt}}

\end{figure}

\begin{table*}[htbp]
\begin{threeparttable}

\caption{Summary of the theoretical SN yield predictions tested in this work. \label{tab:snmodel}}

\centering

\begin{tabular}{lllc}\\ \hline\hline
 label & SN type & description & reference \\ \hline
 W7 & massive Ia & 1D simulation, pure deflagration & \citet{LN18} \\
 LN18p & massive Ia & 2D simulation, pure deflagration & \citet{LN18} \\
 LN18-Lp & massive Ia & 2D simulation, pure deflagration, low core density& \citet{LN18} \\
 LN18-Hp & massive Ia & 2D simulation, pure deflagration, high core density & \citet{LN18} \\
 F14 & massive Ia & 3D simulation, pure deflagration & \citet{Fink14} \\
 F14-L & massive Ia & 3D simulation, pure deflagration, low core density & \citet{Fink14} \\
 F14-H & massive Ia & 3D simulation, pure deflagration, high core density & \citet{Fink14} \\
 WDD2 & massive Ia & 1D simulation, delayed detonation & \citet{LN18} \\
 LN18 & massive Ia & 2D simulation, delayed detonation & \citet{LN18} \\
 LN18-L & massive Ia & 2D simulation, delayed detonation, low core density& \citet{LN18} \\
 LN18-H & massive Ia & 2D simulation, delayed detonation, high core density & \citet{LN18} \\
 S13 & massive Ia & 3D simulation, delayed detonation & \citet{Seitenzahl13b} \\
 S13-L & massive Ia & 3D simulation, delayed detonation, low core density & \citet{Seitenzahl13b} \\
 S13-H & massive Ia & 3D simulation, delayed detonation, high core density & \citet{Seitenzahl13b} \\
 Pak12 & light Ia & double degenerate, violent merger & \citet{Pakmor12} \\
 Sh18 & light Ia & double degenerate, double detonation & \citet{Shen18} \\
 LN20 & light Ia & single degenerate, double detonation & \citet{LN20} \\
 N13 & CC & no rotation  & \citet{Nomoto13} \\
 Suk16 & CC & no rotation, neutrino transport included & \citet{Sukhbold16} \\
 LC18 & CC & no rotation, massive stars $> 25M_\odot$ do not undergo SNe & \citet{LC18} \\ \hline
\end{tabular}

\end{threeparttable}
\end{table*}

\begin{figure}
\begin{center}

\includegraphics[width=\columnwidth]{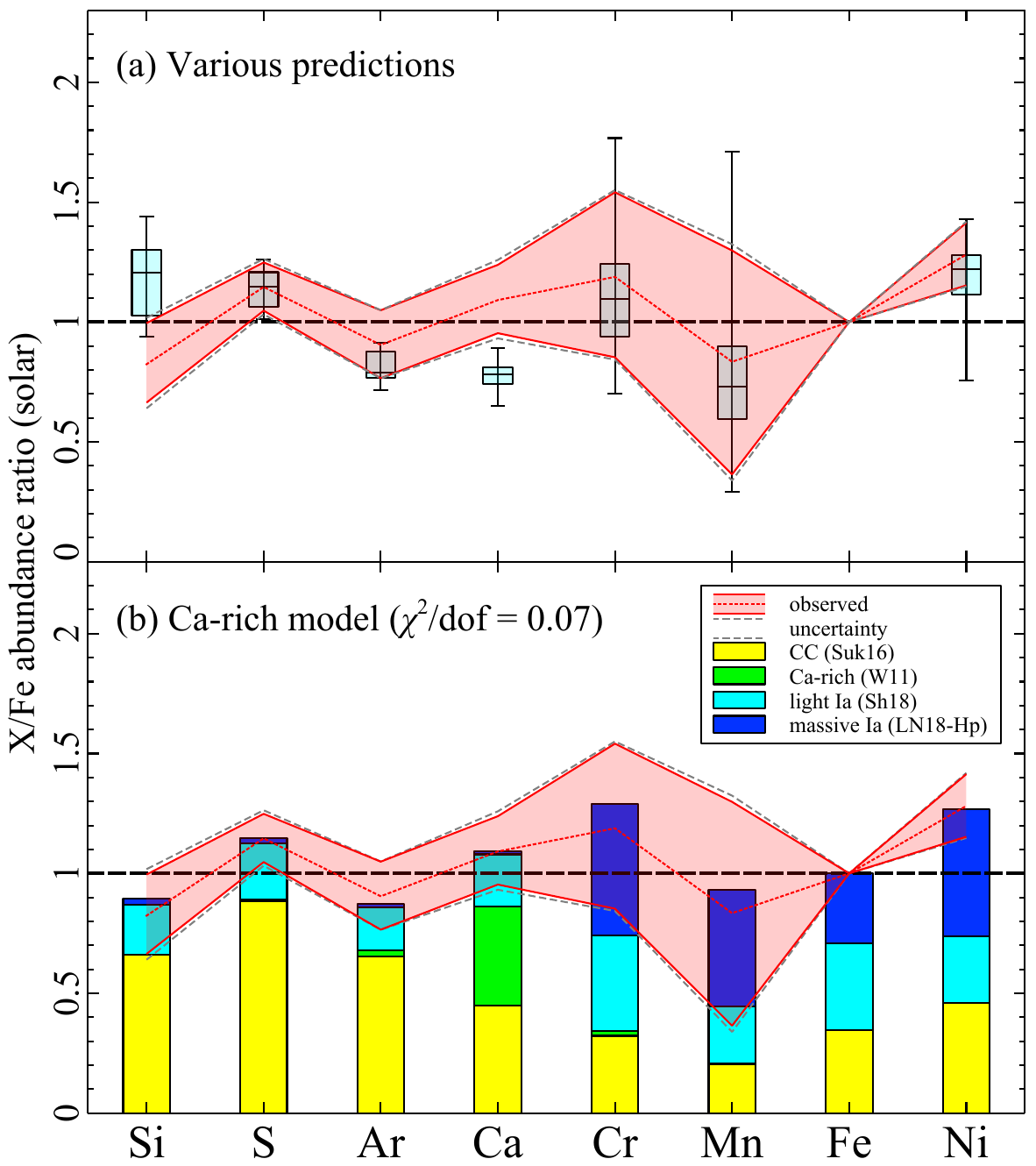}

\end{center}

\caption{(a) Abundance patterns predicted by 126 combinations of SN yields compared to the observed values.
Crosses are predicted abundances from each combination model, and their medians, 1/4, and 3/4-quadrants
are given by boxes. Bars show the max-min intervals of each expected value.
(b) The same as figure\,\ref{fig:patt}(a), but the yield of Ca-rich transient is added (\citealt{Waldman11}, W11).
\label{fig:ca}}

\end{figure}

\subsection{Solar chemical composition of the Ophiuchus core \label{subsec:pattern}}

The Fe abundance about $0.6$\,solar at the Ophiuchus centre ($\lesssim 100$\,kpc)
is consistent with the Suzaku result of \citet{Fujita08},
and possibly those of Chandra and XMM-Newton \citep[][]{Million10a, Werner16, Liu19a, Gatuzz23c},
except for their proposed abrupt peak and/or depletion at the very core ($\lesssim 20$\,kpc).
Then, we show the X/Fe abundance pattern relative to the proto-solar value
in the ICM of the Ophiuchus core (figure\,\ref{fig:patt}(a)).
The abundance ratios are remarkably consistent with the solar value
while the Ni/Fe ratio is relatively high at about $1.3$\,solar,
which is the first robust constraint of Ni abundance in Ophiuchus.
Recently, \citet{Gatuzz23c} reported that the Si/Fe, S/Fe, Ar/Fe, and Ca/Fe values are close to solar ratios
within the 100\,kpc central region of Ophiuchus, which is entirely consistent with our results.

Given that the absolute abundance of Fe is subsolar, the metallicity and composition of the Ophiuchus core
are similar to those of Perseus \citep[][]{Hitomi17, Simionescu19}.
\citet{Mernier18c} and \citet{Simionescu19} discussed in detail
the implications of the observed solar composition in the ICM.
Unless the elemental abundances are effectively \textit{frozen out}
to around the solar value after star formation at redshift\,$= 2$--$3$,
irrespective of subsequent enrichment processes, a plausible interpretation is as follows:
the super-solar ratios of Si/Fe, S/Fe, Ar/Fe, and Ca/Fe by early CCSN enrichment diminished over time
as a substantial quantity of SNe~Ia has contaminated interstellar space with Fe.
Such a trend of chemical evolution is also implied by observations of stellar metallicity in the Milky Way,
where [$\alpha$/Fe] approaches the solar value as [Fe/H] gets close to the solar abundance
\citep[e.g.,][]{Bensby14, Hawkins16, Nandakumar25}.
Nonetheless, we will discuss the averaged properties of SNe integrated over time in Section\,\ref{subsec:sn}.

Regarding only the X/Fe values, it is worth stressing that the solar abundance ratio
is also observed in metal-richer systems (e.g., Centaurus, \citealt{Fukushima22}; \citealt{Mernier25}).
In this sense, \citet{Mernier18b} and \citet{Fukushima23a} performed systematic analyses
and reported a ubiquitous solar chemical composition of the X-ray-emitting haloes in relatively hot systems.
Our abundance pattern will be added as an additional example to support this picture.
Importantly, previous systematic studies have suffered from large uncertainties in abundance measurement (20--40\%),
including differences among samples, in exchange for long integrated exposures
($4.5$\,Ms, \citealt{Mernier18b}; $0.8$\,Ms, \citealt{Fukushima23a}).
The limited energy resolution of CCD instruments and the poor reconstruction of dispersive spectra
from extended sources, for instance, would also contribute to significant uncertainties.
We demonstrated that the Resolve data enable us to estimate abundances with 10--20\% accuracy
among each metal, excluding Cr and Mn, with a single observation of only $217$\,ks.
This fact suggests a decisive advantage in measuring the chemical abundance of the hot ICM,
and will ensure the presence of XRISM in future spectroscopic observations of galaxy clusters.

\subsection{Enrichment by SN explosions \label{subsec:sn}}

We consider a linear combination of the calculations of SN nucleosynthesis
to reproduce the observed abundance pattern from Si/Fe to Ni/Fe.
We adopt calculated models for SN~Ia of \citet{Seitenzahl13b}, \citet{Fink14}, and \citet{LN18}
for SNe from a massive white dwarf (about $1.4 M_\odot$)
and \citet{Pakmor12}, \citet{Shen18}, and \citet{LN20} for those of light progenitors (about $0.8 M_\odot$).
For CCSNe from various progenitor masses, three yields \citep[][]{Nomoto13, Sukhbold16, LC18} are adopted,
each of which is averaged over a Salpeter initial mass function \citep[][]{Salpeter55}.
The models used in this work are summarised in table\,\ref{tab:snmodel},
and we fit the observed pattern with 126 SN combination models.
When fitting the pattern and calculating fit statistics, modelling uncertainties
are added in quadrature to statistical errors for each X/Fe value.

The best-fitting result given by the combination of the F14, Sh18,  and Suk16 models
is presented in figure\,\ref{fig:patt}(a), which yields a good fit with $\chi^2$/dof\,$= 1.1$.
This combination model suggests that the number ratio of SN~Ia to total SNe is about
23\% (massive, 7\%; light 16\%) and SNe~Ia produce 67\% of Fe in the ICM.
A previous study of ICM enrichment by SNe in Ophiuchus estimated
the SN~Ia fraction to be 10--20\% at $\lesssim 100$\,kpc \citep{Gatuzz23c},
which is comparable to our estimate.
To reveal the census of SNe~Ia predicted by current models,
we also plot the contribution of SNe to Fe against SNe~Ia/SNe for all combinations (figure\,\ref{fig:patt}(b)).
While considering various theoretical assumptions of each model (table\,\ref{tab:snmodel}),
the better-fitting models ($\chi^2$/dof\,$< 2$) illustrate a similar enrichment scenario
with a moderate fraction ($21 \pm 5$\%) of SNe~Ia producing the bulk of Fe ($66 \pm 5$\%) in the ICM,
where $52 \pm 23$\% of them come from massive progenitors.
Such a population of SNe~Ia is globally consistent with previous studies of galaxy clusters
\citep[e.g.,][]{Sato07, Simionescu15, Ezer17, Simionescu19, Erdim21}.

Interestingly, some combinations provide a marginal fraction of SNe~Ia below 10\%
despite large $\chi^2$/dof values (lower left area in figure\,\ref{fig:patt}(b)),
wherein the CCSN model is assumed to be LC18.
In the LC18 model, \citet{LC18} assumes that massive stars $> 25M_\odot$ do not undergo
an SN explosion and contribute to enrichment via the stellar mass-loss channel only.
The lack of massive CCSNe leads to relatively low Ar/Fe and Ca/Fe ratios about $1$\,solar compared to other models,
and mixing with SNe~Ia worsens the fit, as these abundances are reduced to subsolar values.
Unlike short-timescale systems subject to ongoing star formation activity,
such as starburst galaxies (Arp~299, \citealt{Mao21}; M82, \citealt{Fukushima24a}),
the ICM of galaxy clusters with a long enrichment history
may require a significant contribution from a massive stellar population.

\subsection{Implications into the ``Ca conundrum'' \label{subsec:carich}}

Figure\,\ref{fig:ca}(a) illustrates the graphical implications of how well recent models predict each X/Fe abundance ratio.
At first glance, the Ca/Fe ratios are uniformly underestimated compared to the observed value, that is, to the solar composition.
Such a \textit{hypocalcemia} problem of nucleosynthesis models, known as the ``Ca conundrum'',
has been mentioned in many previous studies \citep[e.g.,][]{dePlaa07, Werner08, Simionescu19, Fukushima22}.
One solution is proposed by \citet{Mulchaey14} and \citet{Mernier16b},
where the Ca enrichment of the ICM can be attributed not only to normal SNe~Ia but also to Ca-rich gap transients
\citep{Filippenko03, Perets10, Waldman11}.
Both studies estimated that this subclass accounts for about 20--30\% of total SN~Ia events.

We also include the predicted yield of Ca-rich events (W11 from \citealt{Waldman11}),
and the combination model with LN18-Hp, Sh18, and Suk16
significantly improves the fit as shown in figure\,\ref{fig:ca}(b).
In this model, we estimate the fraction of Ca-rich gap transients to total SNe~Ia is 21\%,
where other good combinations ($\chi^2$/dof\,$< 2$) yield 13--33\% fraction.
Our estimation is consistent with the estimation by \citet{Mulchaey14} and \citet{Mernier16b}.
While the fraction of galactic Ca-rich events had been estimated to be $< 20$\% \citep[e.g.,][]{Perets10, Li11},
more updated observational constraints allow more frequent events $> 30$\%
at larger distances from galaxies \citep{Frohmaier18}.
Given that the abundance pattern of the ICM may be more consistent
with those of the outer edge of galaxies than of their inner region, dominated by the member stellar population,
Ca-rich transients can naturally solve the Ca conundrum.

\section{Conclusions \label{sec:concl}}

We have analysed the Resolve data of the Ophiuchus cluster core and measured the metal abundance pattern.
The results are summarised as follows.

\begin{itemize}

\item[1.] We find that the abundance pattern of the Ophiuchus core is consistent with solar,
which is reminiscent of the Hitomi constraint of Perseus \citep[][]{Simionescu19}.
The uncertainties of abundance measurement are 10--20\%,
less than in previous systematic studies using large samples \citep[e.g.,][]{Mernier18b},
demonstrating an excellent capability of XRISM for elemental abundance studies.

\item[2.] Recent SN nucleosynthesis models prefer a specific chemical enrichment history
that constitutes 15--25\% of total SNe being SN~Ia, and 60--70\% of Fe in the ICM being forged by them.
The CCSN contribution from massive stars with $> 25M_\odot$ progenitors
may be crucial to reproducing the observed abundances accurately,
which differs from the case of starburst systems \citep[e.g.,][]{Mao21}.

\item[3.] The observed Ca/Fe ratio about $1$\,solar is hard to explain
by the standard SN combination model.
The Ca-rich gap transients, thermonuclear events classified as a subclass of SNe~Ia, can be one solution
with a substantial contribution of about 13--33\% of SNe~Ia, which is in plausible agreement
with the pictures provided by direct SN studies \citep[e.g.,][]{Frohmaier18}.

\end{itemize}




\begin{ack}
The authors are grateful to the anonymous referee for their helpful comments and suggestions,
improving our manuscript drastically.
We appreciate Dr.~Y.~Kanemaru for providing excellent Xtend images.
We also thank the project members who are working on the mission and science operations of XRISM.
The Japan Aerospace Exploration Agency (JAXA), the National Aeronautics and Space Administration (NASA),
and the European Space Agency (ESA) have collaborated to develop and operate XRISM internationally.
In addition to the three space agencies, universities and research institutes from Japan, the United States,
and European states have contributed to the development of satellites, science instruments,
and data-processing software, as well as to the further preparation of scientific observation plans.
This work is also based on XMM-Newton, the ESA science mission
with instruments and contributions directly funded by ESA Member States and NASA of the United States.
The figures in this article were generated using \textsc{veusz} (\url{https://veusz.github.io/}) and \textsc{Python} version 3.10.
\end{ack}

\section*{Funding \label{sec:fund}}
This work was supported by JSPS KAKENHI grant Nos. 22H00158, 23H04899, and 25H00672 (YF).

\section*{Data availability \label{sec:data}}
The XRISM data set analysed in this article is for exclusive use until April 16, 2026;
afterwards, the data will be made available publicly via the XRISM data archive
\footnote{$\langle$\url{https://data.darts.isas.jaxa.jp/pub/xrism/data/obs/}$\rangle$}
\footnote{$\langle$\url{https://heasarc.gsfc.nasa.gov/FTP/xrism/data/obs/}$\rangle$}.
The XMM-Newton Science Archive\footnote{$\langle$\url{https://nxsa.esac.esa.int/nxsa-web/}$\rangle$}
stores and distributes the data set analysed in the Appendix section.

\appendix

\section{O, Ne, and Mg abundances \label{sec:xmm}}

\begin{figure}
\begin{center}

\includegraphics[width=\columnwidth]{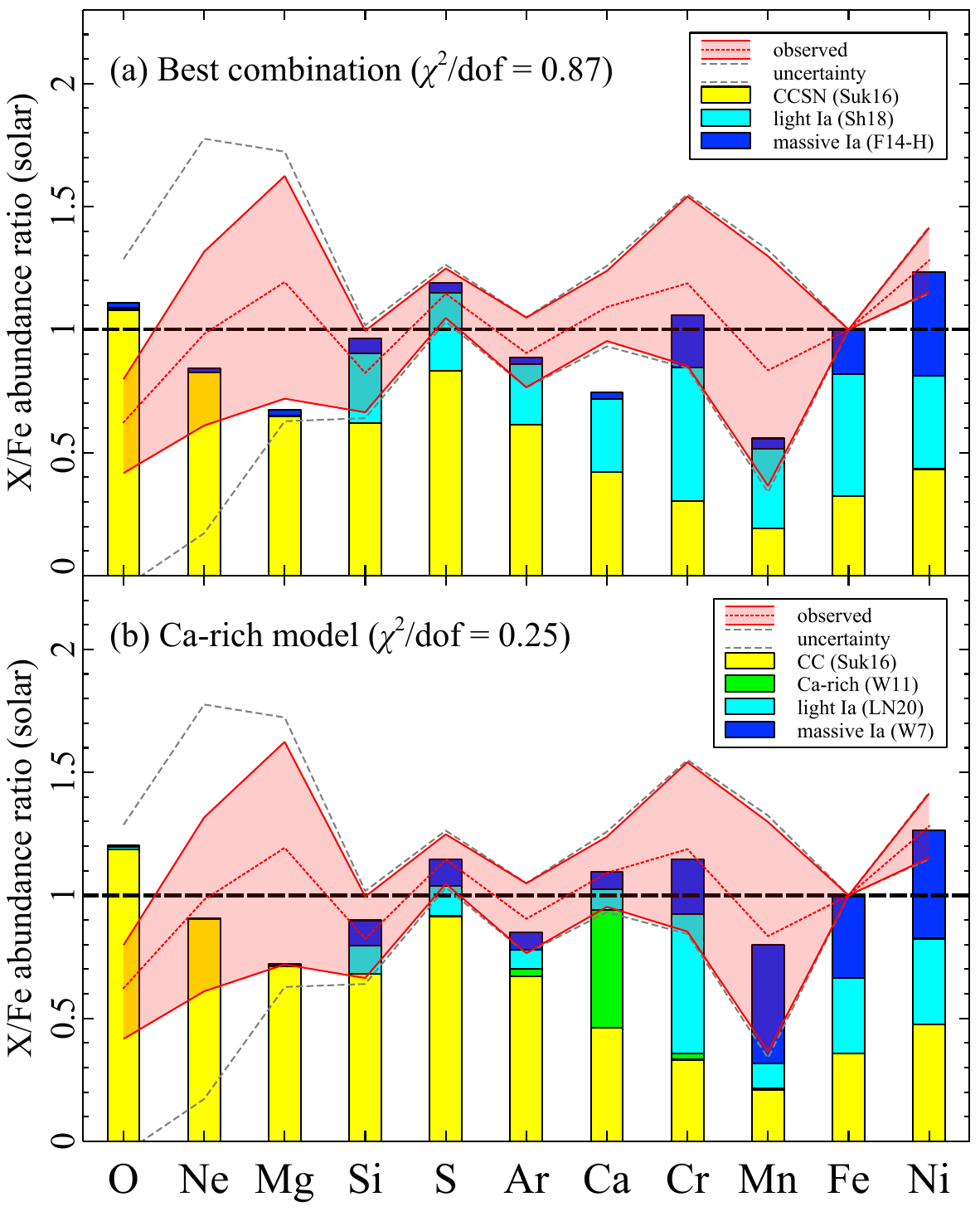}

\end{center}

\caption{(a) The same as figure\,\ref{fig:patt}(a), but including the O/Fe, Ne/Fe, and Mg/Fe ratios.
(b) The same as figure\,\ref{fig:ca}(b) with the O/Fe, Ne/Fe, and Mg/Fe ratios.
\label{fig:xmm}}

\end{figure}

In this Appendix, we present the O, Ne, and Mg abundances in our abundance pattern.
We derive these abundances using the XMM-Newton archival data set ($\texttt{OBSID} = \texttt{0505150101}$),
pointing the centre of the Ophiuchus cluster.
To compare with the Resolve results with a modest spatial resolution,
we extract spectra from the $2.5$\,arcmin core from MOS \citep[][]{Turner01} and pn \cite[][]{Strueder01},
covering the entire Resolve FoV and its surrounds.
Similar to the analysis procedures described in Sections\,\ref{subsec:fit} and \ref{subsec:abund},
we employ the 1T, 2T, and DEM models in both the $0.4$--$9$\,keV and $0.4$--$2$\,keV bands
and estimate the  O, Ne, Mg, and Fe abundances.
The elements heavier than Al are linked to Fe, and the others follow the same setting in Section\,\ref{subsec:fit}.

The Fe abundance is well constrained around $0.6$--$0.7$\,solar,
and the O, Ne, and Mg ones show large differences of up to 50\%
between the broad and local band fits across all modelling.
What causes this large discrepancy may simply be the limitations of the energy resolution of the CCD instruments
and/or the relatively high temperature of the ICM ($kT \sim 8$--$9$\,keV),
as well as possible different metal content of the low-temperature components responsible for these light metals
from that of the high-temperature components accessible with the Resolve spectrum.
\citet{Mernier25} discusses such a plausible multi-metallicity ICM
in Centaurus with a stronger cool core than that of Ophiuchus.
Nevertheless, we adopt the results from the DEM model in the $0.4$--$2$\,keV band:
$\textup{O/Fe} = 0.6 \pm 0.2 (\pm 0.6)$\,solar, $\textup{Mg/Fe} = 0.9 \pm 0.4 (\pm 0.7)$\,solar,
and $\textup{Ne/Fe} = 1.2 \pm 0.5 (\pm 0.6)$\,solar.

Figures\,\ref{fig:xmm}(a) and (b) demonstrate the abundance pattern
(O/Fe, Ne/Fe, Mg/Fe, Si/Fe, S/Fe, Ar/Fe, Ca/Fe, Cr/Fe, Mn/Fe, and Ni/Fe)
and the best-fitting SN combination models.
With the light $\alpha$-element abundances, despite large systematic errors,
the preferred enrichment scenario of better combinations yields $9 \pm 7$\% and $9 \pm 5$\%
for massive and light SNe~Ia, respectively, and $82 \pm 7$\% for CCSNe.
As well, the Ca-rich transient contribution improves the fit with a fraction $24 \pm 5$\% of SNe~Ia.
These results are completely consistent with the main conclusions in Sections\,\ref{subsec:sn} and \ref{subsec:carich}.
The crucial advantage of including the contribution from the Ca-rich transient channel is also discussed in \citet{Sarkar25b},
focusing on the strong cool core cluster Abell~2029 with the XRISM data sets.

\bibliographystyle{mnras}
\bibliography{paper_ref}

\end{document}